\documentclass[%
notitlepage,
superscriptaddress,
 amsmath,amssymb,
 aps,
prb]{revtex4-1}
\usepackage{bm,amsmath,amssymb}
\usepackage{graphicx}
\usepackage{hyperref}

\begin{document}

\title{Anisotropic magnetic responses of topological crystalline superconductors}

\author{Yuansen Xiong}
\affiliation{Department of Applied Physics, Nagoya University, Nagoya 464-8603, Japan}

\author{Ai Yamakage}
\affiliation{Department of Applied Physics, Nagoya University, Nagoya 464-8603, Japan}
\affiliation{Institute for Advanced Research, Nagoya University, Nagoya 464-8601, Japan}

\author{Shingo Kobayashi}
\affiliation{Department of Applied Physics, Nagoya University, Nagoya 464-8603, Japan}
\affiliation{Institute for Advanced Research, Nagoya University, Nagoya 464-8601, Japan}

\author{Masatoshi Sato}
\affiliation{Yukawa Institute for Theoretical Physics, Kyoto University, Kitashirakawa Oiwake-Cho, Kyoto 606-8502, Japan}

\author{Yukio Tanaka}
\affiliation{Department of Applied Physics, Nagoya University, Nagoya 464-8603, Japan}

\date{\today}

\begin{abstract}
 Majorana Kramers pairs emerged on surfaces of time-reversal-invariant topological crystalline superconductors show the Ising anisotropy to an applied magnetic field. 
We clarify that crystalline symmetry uniquely determines the direction of the Majorana Ising spin for given irreducible representations of pair potential, deriving constraints to topological invariants.
Besides, necessary conditions for nontrivial topological invariants protected by the $n$-fold rotational symmetry are shown.
\end{abstract}

\maketitle

\section{Introduction}

Topological superconductors are gapped systems hosting gapless states on their surfaces \cite{hasan10, qi11, tanaka12, ando13} as Andreev bound states \cite{buchholtz81, hara86, hu94, tanaka95}.
The gapless surface states behave as Majorana fermions, which are self-conjugate particles and protected by the topological invariant associated to (broken) symmetries.
Due to the stability and the so-called non-Abelian statistics derived from the self-conjugate property, one would expect that topological superconductors can be a platform of the fault-tolerant topological quantum computation \cite{nayak08}.

For further achievement, it is necessary to detect and manipulate the Majorana fermions.
Applying an external magnetic field is a one way to destruct Majorana fermions in time-reversal-invariant (DIII \cite{schnyder08, kitaev09, ryu10}) topological superconductors since a magnetic field breaks time-reversal symmetry.
However,
Majorana fermions still remain gapless when an applied magnetic field is normal to a certain direction.
Namely, Majorana fermions exhibit the Ising anisotropy \cite{sato09, chung09, nagato09, volovik10, shindou10, silaev11, yada11, mizushima11, mizushima12, mizushima13}.
In this sense, Majorana fermions are referred to as Majorana Ising spin.

Shiozaki and Sato \cite{shiozaki14} have unveiled that the underlying mechanism of Majorana Ising spin is the protection by crystalline symmetry, i.e., an extension from topological crystalline superconductivity \cite{chiu13, morimoto13, ando15, chiu15}.
In this paper, we develop the theory and find that the direction of Majorana Ising spin is uniquely determined for a given irreducible representation \cite{sigrist91} of the pair potential. 
The obtained result can be applied to all the space groups hence we believe that it is useful for studies on topological-superconductor materials and experiments.

The paper is organized as follows.
We start with a brief review on Majorana Ising spin in Sec. \ref{preliminary} and clearly summarize issues to be addressed.
We derive conditions for nontrivial topological invariant in Sec. \ref{tiir} in systems with time-reversal and crystalline symmetries.
From the obtained conditions, one finds the direction of Majorana Ising spin and summarizes it in tables (Appendix \ref{tables}).
Besides Majorana Ising spin, in Sec. \ref{n-fold},  we also derive the winding number corresponding to the surface Majorana fermions protected by $n$-fold rotational symmetry, in a manner similar to that in Sec. \ref{tiir}.
An example of the application of our general theory is shown in Sec. \ref{examples}.
We finally summarize the paper in Sec. \ref{conclusion}.

\section{Preliminary}
\label{preliminary}
Before going into the main discussion, we first review zero-energy states and the associated topological invariants in superconductors.
A BdG Hamiltonian $H(\bm k)$ has the form
\begin{align}
 H(\bm k)  
 &= \begin{pmatrix}
  h(\bm k) - \mu & \Delta(\bm k)
  \\
  \Delta(\bm k) & -h(\bm k) + \mu
 \end{pmatrix} 
= \left[h(\bm k)-\mu \right] \tau_z + \Delta(\bm k) \tau_x,
\end{align}
in the basis of $(c_{\uparrow}(\bm k), c_{\downarrow}(\bm k), c^\dag_{\downarrow}(-\bm k), -c^\dag_{\uparrow}(-\bm k))$, where $\uparrow$ and $\downarrow$ denote the spin up and down, respectively, and the spin indices in $h(\bm k)$ and $\Delta(\bm k)$ are implicit.
Note that one can choose $\Delta(\bm k) = \Delta(\bm k)^\dag$ for time-reversal-invariant superconductors.
The Hamiltonian preserves time-reversal $T$ symmetry
\begin{align}
 h(\bm k) &= T^{-1} h(-\bm k) T,
 \
 \Delta(\bm k) = T^{-1} \Delta(-\bm k) T,
\
 H(\bm k) = \mathcal T^{-1} H(-\bm k) \mathcal T,
 \
 \mathcal T = \tau_0 T,
\end{align}
and particle-hole $\mathcal C$ symmetry
\begin{align}
 H(\bm k) = -\mathcal C^{-1} H(-\bm k) \mathcal C,
 \
 \mathcal C = \tau_y \mathcal T.
\end{align}
Combining these symmetries, chiral symmetry holds;
\begin{align}
  \{ \Gamma, H(\bm k) \} = 0,
  \
  \Gamma = \mathcal C \mathcal T = \tau_y.
  \label{Gamma}
\end{align}

\begin{figure}
\centering
\includegraphics{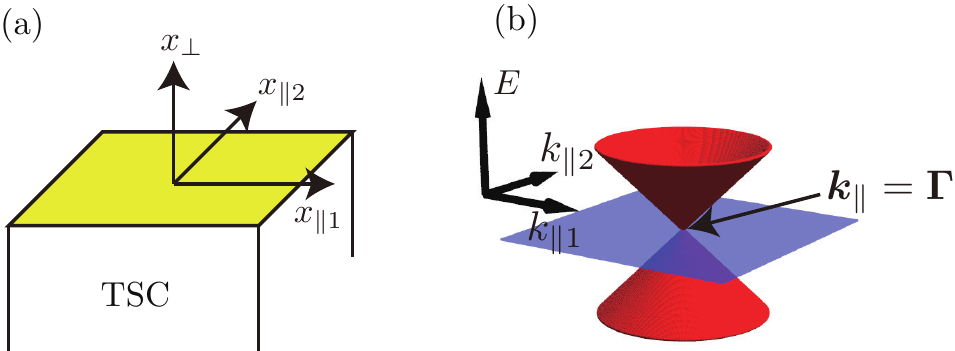}
\caption{(a) Topological superconductor (TSC) with the surface on $x_\perp = 0$. (b) Energy dispersion of the surface Majorana zero modes located at the time-reversal-invariant momentum $\bm k_{\parallel} = \bm\Gamma$.}
\label{surface}
\end{figure}

Next, we introduce the topological invariant corresponding to the number of zero-energy states on the surface, which are located on $x_\perp = 0$ [Fig. \ref{surface}(a)].
%
The time-reversal-invariant momentum at which the zero-energy states appear is set to $\bm k_\parallel = \bm \Gamma$ [Fig. \ref{surface}(b)].
The one-dimensional topological invariant $W$ \cite{sato-tanaka10}  is given by
\begin{align}
  W = \frac{i}{4\pi}
  \int_{-\pi/a_\perp}^{\pi/a_\perp} d k_\perp \mathrm{tr}
  \left[ \Gamma
    H(k_\perp)^{-1} \frac{\partial H(k_\perp)}{\partial k_\perp}
  \right]_{\bm k_\parallel = \bm \Gamma} \in \mathbb Z,
  \
  H(k_\perp) = H(\bm k)|_{\bm k_\parallel = \bm \Gamma},
\end{align}
where $H(k_\perp) = H(k_\perp + 2 \pi/a_\perp)$.
This invariant is equal to the number of the zero-energy surface states (see Appendix \ref{bec}).
In time-reversal-invariant spinful  systems, however, the above topological invariant always vanishes owing to time-reversal symmetry \cite{schnyder08, kitaev09, ryu10, kobayashi-tanaka16},
which requires $\{ \mathcal T, \Gamma \} = 0$ and
\begin{align}
 W^* &= \frac{-i}{4\pi}  
 \int_{-\pi/a_\perp}^{\pi/a_\perp} d k_\perp
 \mathrm{tr}
 \left[
	\Gamma^* H(k_\perp)^{* -1} \frac{\partial H(k_\perp)^*}{\partial k_\perp}
 \right] 
 \nonumber\\&
 =
 \frac{-i}{4\pi}  
 \int_{-\pi/a_\perp}^{\pi/a_\perp} d k_\perp
 \mathrm{tr}
 \left[
	(-\Gamma) H(-k_\perp)^{-1} \frac{\partial H(-k_\perp)}{\partial k_\perp}
 \right] 
= -W,
 \label{zeroW}
\end{align}
using $\mathcal T H(\bm k) \mathcal T^{-1} = H(-\bm k)$, hence $W=0$.

The topological invariant can take a finite value with the help of an order-2 symmetry operation $U$ that involves the spin and respects the surface: $[U, H(k_\perp)]=0$. 
Now we introduce a modified chiral operator $\Gamma_U$ as
\begin{align}
 \Gamma_U = e^{i \phi_U} U \Gamma, 
 \
 \Gamma_U^2 = 1,
\end{align}
where the phase $\phi_U$ is chosen to satisfy $\Gamma_U^2 =1$.
The modified topological invariant $W[U]$ is given by replacing $\Gamma$ with $\Gamma_U$;
\begin{align}
 W[U] = \frac{i}{4\pi} \int_{-\pi/a_\perp}^{\pi/a_\perp} d k_\perp
 \mathrm{tr}
 \left[
	\Gamma_U H(k_\perp)^{-1} \frac{\partial H(k_\perp)}{\partial k_\perp}
  \right],
\end{align}
which is free from the condition of Eq. (\ref{zeroW}) when the following condition is satisfied;
\begin{align}
 [\mathcal  T, \Gamma_U ] = 0.
 \label{TGU}
\end{align}
Normally, order-2 symmetry operations stem from crystalline point/space-group symmetries such as two-fold rotations and reflections with respect to the $x_\perp$ axis.
This means that systems with $W[U] \ne 0$ are interpreted as a one-dimensional topological crystalline superconductor.

In the last part of this section, we review that $W[U]$ naturally explains the Ising-anisotropic response to a magnetic field \cite{mizushima12,shiozaki14}.
The symmetry operation $U$ is taken to be a two-fold rotation or a reflection.
The symmetry operation $U$ flips or keeps the direction of applied magnetic field, i.e., $\{ U, H_{\rm mag} \} = 0$ or $[U, H_{\rm mag}] = 0$, respectively. 
Here $H_{\rm mag}$ denotes the Hamiltonian of magnetic field including the Zeeman and vector potential terms.
These operations are summarized in Table \ref{sym_mag}.
\begin{table}
\centering
\caption{Symmetry of magnetic field $\bm B$ applied along the $x_\perp$, $x_{\parallel 1}$, and $x_{\parallel 2}$ directions, which are depicted in Fig. \ref{surface}. $C_2(x_\perp)$ is the two-fold rotation along the $x_\perp$ axis. $\sigma(x_i x_j)$ is the mirror reflection with respect to the $x_i x_j$ plane. These are symmetry operations of the semi-infinite system with the surface of $x_\perp = 0$. $-$ ($+$) indicates that the magnetic field is (not) flipped by the symmetry operation. $\bm S$ denotes the direction of Majorana Ising spin protected by the topological invariant $W[U]$ for $U=C_2(x_\perp)$, $\sigma(x_\perp x_{\parallel 1})$, and $\sigma(x_\perp x_{\parallel 2})$.}
\begin{tabular}{c|ccc|c}
 \hline\hline
 $U$ & $\bm B \parallel \bm x_\perp$ & $\bm B \parallel \bm {x}_{\parallel 1}$ & $\bm B \parallel \bm x_{\parallel 2}$ & $\bm S$
 \\ \hline
 $C_2(x_\perp)$ & $+$ & $-$ & $-$ & $x_\perp$
 \\
 $\sigma(x_\perp x_{\parallel 1})$ & $-$ & $-$ & $+$ & $x_{\parallel 2}$
 \\
 $\sigma(x_\perp x_{\parallel 2})$ & $-$ & $+$ & $-$ & $x_{\parallel 1}$
 \\ \hline\hline
\end{tabular}
\label{sym_mag}
\end{table}
In the former case ($\{ U, H_{\rm mag} \} = 0$), since the modified chiral symmetry still remains $\{ \Gamma_U, H_0 + H_{\rm mag} \} = 0$, the modified topological invariant $W[U]$ is well-defined and shares the same number as that in the absence of magnetic field as long as $H_{\rm mag}$ is small enough, while not in the latter case ($[U, H_{\rm mag}] = 0$).
From Table \ref{sym_mag}, the latter case is realized only in the case that magnetic field is applied for a specific direction in each symmetry operation.
Therefore, the zero-energy surface states protected by $W[U]$ are annihilated only by the magnetic field along the specific direction. 
Namely, Majorana fermions on the surface acts as an Ising spin under a magnetic field.

In the following sections, developing the theory, we show only one winding number among $W[C_2(x_\perp)]$, $W[\sigma(x_\perp x_{\parallel 1})]$, and $W[\sigma(x_\perp x_{\parallel 2})]$ is possible to take a finite value for a given surface and an irreducible representation of pair potential, i.e., the anisotropy of magnetic response is uniquely determined, irrespective of the details of the system.

\section{Topological invariants for irreducible representations}
\label{tiir}

We show that only one among three possible topological invariants $W[{C_2(x_\perp)}]$, $W[{\sigma(x_\perp x_{\parallel 1})]}$, and $W[{\sigma(x_\perp x_{\parallel 2})}]$ can become finite in a given superconducting pair potential.
This is the decisive evidence of Majorana Ising spin.

\subsection{Symmetry of crystalline systems including a surface}

\begin{figure}
\centering
\includegraphics[scale=0.5]{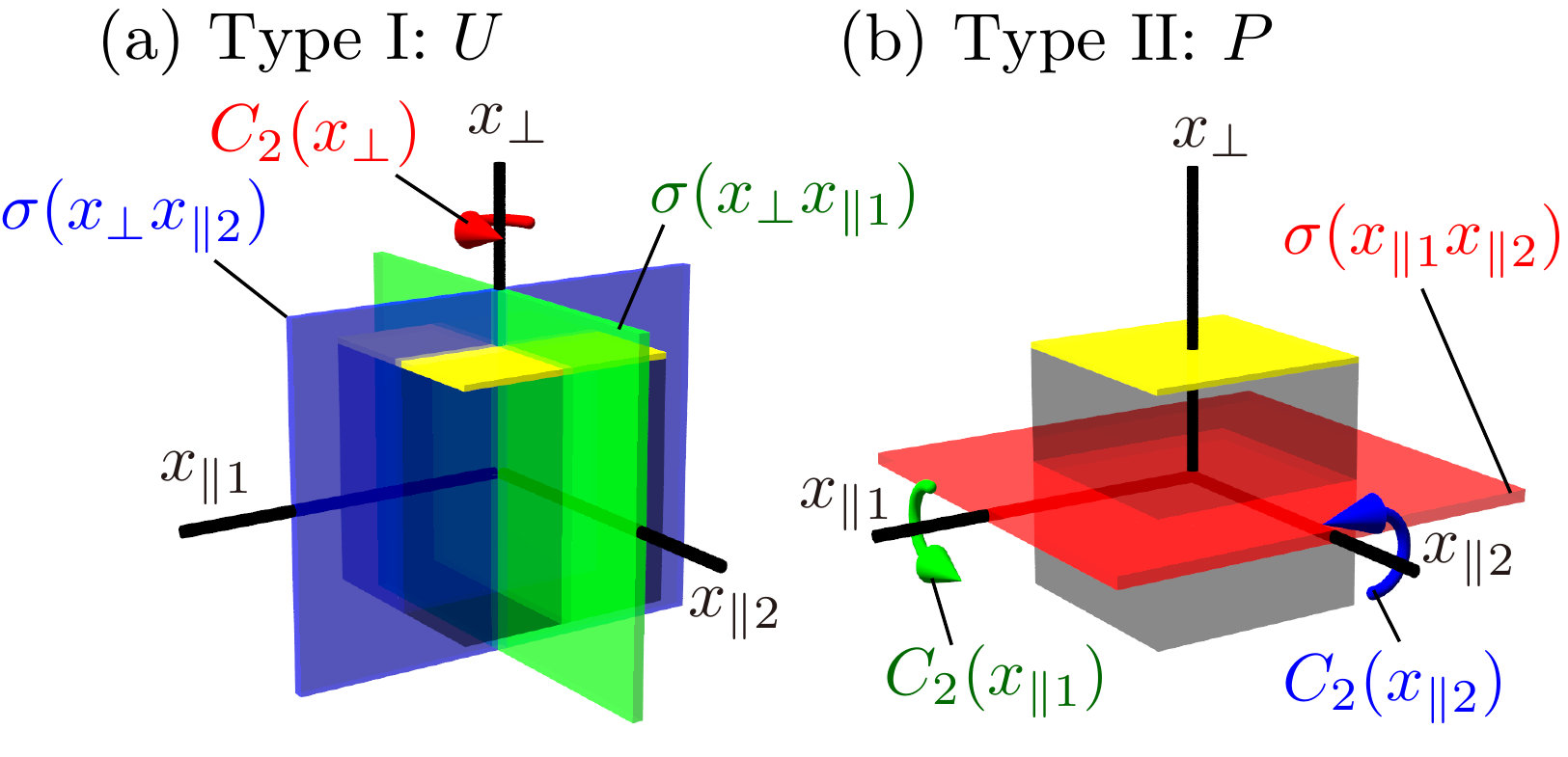}
\caption{Point-group symmetry operations which preserve (a) and invert (b) the surface  of $x_\perp = 0$.}
\label{symz}
\end{figure}

In crystalline systems, all the symmetry operations other than time reversal and particle-hole transformation are elements of a space group.
Here we focus on the momentum line including the time-reversal-invariant momentum $\bm k = \bm \Gamma$ along which the one-dimensional topological invariant is defined.
Symmetry operations respecting the $\bm k = \bm \Gamma$ point are (screw) rotation and (glide) reflection, which are classified into those preserving (type-I) and inverting (type-II) the surface of $x_\perp = 0$.
The type-I symmetry operations are two-fold (screw) rotation $C_2(x_\perp)$ along the $x_\perp$ axis and mirror (glide) reflections $\sigma(x_\perp x_{\parallel 1})$ with respect to $x_{\perp} x_{\parallel 1}$ plane and $\sigma(x_\perp x_{\parallel 2})$ with respect to $x_{\perp} x_{\parallel 2}$ plane [Fig. \ref{symz}(a)].
The type-II symmetry operations, on the other hand, are two (screw) rotations [$C_2(x_{\parallel 1})$ and $C_2(x_{\parallel 2})$] and one mirror (glide) reflection [$\sigma(x_{\parallel 1} x_{\parallel 2})$], as shown in Fig. \ref{symz}(b).
Afterwards, we denote a type-I operation by $U_i$ then we have
\begin{align}
  [U_i, h(k_\perp)] = 0.
\end{align} 
The spatial inversion $I$ is represented in terms of $U_i$ as
\begin{align}
 I = U_i P_i,  
\end{align}
for $i=1,2,3$, where $P_i$ is a type-II symmetry operation, i.e., 
\begin{align}
 h(k_\perp) = P_i^\dag h(-k_\perp) P_i.
\end{align}

\subsection{Symmetry operations in superconducting states}

Now define symmetry operations in a superconductor.
A superconductor keeps a crystal symmetry $S$ ($U_i$ or $P_i$) where the pair potential $\Delta (\bm k)$ is a one-dimensional representation of $S$;
\begin{align}
 \Delta(\bm k) = \chi(S) S^\dag \Delta(\bm k') S, 
\end{align}
where $\chi(S)$ is the character of the one-dimensional representation and $\bm k'$ is the momentum transformed by $S$.
Then, 
the symmetry operation $\tilde S$ in the superconducting state is defined as
\begin{align}
 \tilde S = 
 \begin{pmatrix}
   S & 0
   \\
   0 & \chi(S) S
 \end{pmatrix},
\end{align} 
which satisfies $\tilde S^{\dag} H(\bm k) \tilde S =H(\bm k')$.
If $S$ obeys $S^2 \propto 1$, $\chi (S)$ of the one-dimensional representation is either $\chi (S) =1 $ or $\chi (S) =-1$, and   
\begin{align}
\tilde S = \begin{cases}
	S \tau_0, & \textrm{for} \ \chi(S)=+1,
	\\
	S \tau_z, & \textrm{for} \ \chi(S) = -1.
 \end{cases}
\end{align}
In these cases, one obtains the following relation 
\begin{align}
 \tilde S^\dag \Gamma \tilde S = \chi(S) \Gamma,
\end{align}
for $\Gamma =\tau _y$ [see Eq. (\ref{Gamma})].

\subsection{Topological invariant}
\label{ti}

In the following, we derive necessary conditions for finite-valued topological invariants, which is defined by
\begin{align}
 W[{\tilde U_i, \bm x_\perp}] = \frac{i}{4 \pi}
 \int_{\bm k \parallel \bm x_\perp} d \bm k  \cdot
 \mathrm{tr} \left[
  \Gamma_{\tilde U_i} H(\bm k)^{-1}
	 \frac{\partial H(\bm k)}{\partial \bm k}
 \right]_{\bm k_\parallel = \bm \Gamma}.
\end{align}
$\bm x_\perp$ indicates the direction normal to the surface, i.e., $W[\tilde U_i, \bm x_\perp]$ is the number of the Majorana zero modes on the surface perpendicular to $\bm x_\perp$.
Note that the type-II symmetries $P_i$ may define an  topological invariant but it does not correspond to the zero-energy surface states since the surface is not invariant against $P_i$. 
This is why only the type-I symmetries $U_i$ are considered here.
Glide reflection along the direction parallel to the surface, e.g., $a$-glide with respect to the $ac$ plane for the $ab$ surface,  is one of the possible type-I $U_i$ symmetries for the winding number.
Screw rotation, however, is not used for the winding number because the surface is not invariant by the operation.
Glide reflection that translates a system along the direction normal to the surface and screw rotation may define a bulk invariant although the bulk-edge correspondence does not hold, as type-II $P_i$ symmetry.
Henceforth, for the rotational symmetries, we suppress the suffix $\bm x_\perp$ as  $W[{\tilde C_2(x_\perp), \bm x_\perp}] = W[{\tilde C_2(x_\perp)}]$, due to the uniqueness of the directions of the integrals, i.e., $\bm x_\perp$ must be along the rotational axis.

Now we derive the constraint to $W[{\tilde U_i, {\bm x}_\perp}]$ by the symmetries.
One gets
\begin{align}
 W[{\tilde U_i, \bm x_\perp}] = +p(\tilde U_i, \tilde U_l) \chi(U_l) W[{\tilde U_i, \bm x_\perp}],
 \
 W[{\tilde U_i, \bm x_\perp}] = -p(\tilde U_i, \tilde P_l) \chi(P_l) W[{\tilde U_i, \bm x_\perp}].
\end{align}
These equations are derived by applying unitary transformations by $\tilde U_l$ and by $\tilde P_l$.
Here we introduce $p(A, B)$ as
\begin{align}
   B^{-1} A B = p(A, B) A.
\end{align}
Note that $\tilde U_l$ includes the $n$-fold rotation (if exist) in addition to the two-fold rotations.
In consequence, the conditions of 
\begin{align}
 p(\tilde U_i, \tilde U_l) \chi (U_l) = -p(\tilde U_i, \tilde P_l) \chi(P_l) = 1,
 \label{condition}
\end{align}
and of $[\mathcal T, \Gamma_{\tilde U_i}] = 0$ [Eq. (\ref{TGU})] are necessary for $W[\tilde U_i, \bm x_\perp] \ne 0$.
From the above condition, $\chi(U_i) = 1$ is derived because of $p(U_i, U_i)=1$.

Next, we prove that the two-fold symmorphic symmetry operations, rotations and reflections, satisfy the condition of Eq. (\ref{TGU}) while the nonsymmorphic ones, glide reflections, do not on the Brillouin zone boundary.
Symmetry operations are represented by the direct product of real-space part $O_{i}$ and spin part $\Sigma_i$, $U_i = O_{i} \Sigma_i$.
For two-fold rotations and mirror reflections, the real-space part $O_{i}$ is an orthogonal matrix with $O_i^2 = 1$ and $[O_i, \mathcal T] = 0$.
The spin part is given by Pauli matrices then $\Sigma^\dag_i = \Sigma_i$, $\Sigma^2_i = 1$, and $\{ \Sigma_i, \mathcal T \} = 0$.
As a result, the chiral operator is given by $\Gamma_{\tilde U_i} = \tilde U_i \tau_y$ so that the condition of Eq. (\ref{TGU}) holds.
For glide reflections, on the contrary, the orbital part $O_{i}$ on the Brillouin zone boundary is purely-imaginary matrix hence the condition Eq. (\ref{TGU}) is not satisfied, i.e., 
\begin{align}
 W[{\tilde U_i, \bm x_\perp}] = 0, 
 \
 \mathrm{for}
 \
 \bm k_\parallel \cdot \bm \tau = \pm \pi/2,
 \label{glide}
\end{align}
with $\tilde U_i$ being a glide reflection, where $\bm \tau$ is the translation vector of the glide reflection (for details, see Appendix \ref{rep}).

For symmorphic space groups, the necessary condition for $W[\tilde U_i, \bm x_\perp] \ne 0$ is easily obtained as follows.
The commutation relations of the representations for symmetry operations in a point group are uniquely determined to be
\begin{align}
 \{ U_i, U_j \} = 
 \{P_i, P_j \} = 
 [U_i, P_i]  = 
 \{ U_i, P_j \}  = 
  0, 
 \
 i \ne j,
\end{align}
in spinful systems.
With the help of the above relation, the condition Eq. (\ref{condition}) reduces to
%
%
%
%
%
%
\begin{align}
 \chi(U_i) = -\chi(U_j) = -\chi(P_i) = \chi(P_j) = 1.
 \label{chis}
\end{align}
Here, $\chi(O)$ is the character of $O$ hence the possible topological invariant is determined only from the representation theory of point group, irrespective of details of the system, as summarized in the tables in Appendix \ref{tables}.
An example for a nonsymmorphic space group is also shown in Appendix \ref{tables}.
The condition of $\chi(U_i) \chi(P_i) = \chi(U_i P_i) = -1$ is extracted from the above equations.
This means that the character of the spatial inversion $I = U_i P_i$ must be $-1$ for the existence of topological superconductivity. 
That is consistent with the absence of time-reversal-invariant Majorana fermion in even-parity superconductors \cite{sato10}.

Finally, we show that two of $W[C_2(x_\perp)]$, $W[\sigma(x_\perp x_{\parallel 1})]$, and $W[\sigma(x_\perp x_{\parallel 2})]$ always vanish.
Here $C_2(x_\perp)$ is the two-fold (not screw) rotation, $\sigma(x_\perp x_{\parallel j})$ is the mirror or glide reflection with respect to the $x_\perp x_{\parallel j}$ plane.
The statement is immediately seen from Eq. (\ref{chis}) for symmorphic space groups: $\chi(U_i)=1$ and $\chi(U_j)=1$ are not simultaneously satisfied.
This is also true at $\bm k_\parallel = \bm 0$ for nonsymmorphic space groups since the commutation relations of symmetry operations are the same as those for the symmorphic space group.
When $\sigma(x_\perp x_{\parallel 1})$ is the $x_{\parallel 1}$-glide reflection, the commutation relation changes from the symmorphic one at the boundary $k_{\parallel 1} = \pi/a_{\parallel 1}$.
$W[\sigma(x_\perp x_\parallel 1)]$, however, vanishes from Eq. (\ref{glide}).
In consequence, 
it is impossible that 
two of $W[C_2(x_\perp)]$, $W[\sigma(x_\perp x_{\parallel 1})]$, and $W[\sigma(x_\perp x_{\parallel 2})]$ simultaneously take nontrivial values.

\section{Winding number protected by \texorpdfstring{$n$}{n}-fold rotational symmetry}
\label{n-fold}

 Besides order-2 symmetries, we clarify the winding number protected by the $n$-fold ($n \geq 3$) rotational $C_n$ symmetry, $[C_n, H(k)]=0$.
 We derive the necessary condition for nonzero topological invariant associated with $C_n$ for spinful systems.
 The spinless case was discussed in Ref. \cite{izumida16}.
 
\subsection{Definition}
$C_n$ is represented by $C_n = e^{-i j_z 2\pi/n}$, where $j_z$ denotes the total angular momentum along the rotational axis.
For spinful systems, $C_n^n = -1$ and the eigenvalue of $C_n$ is obtained to be $e^{-i \mu 2\pi/n}$ for $\mu = 0, \cdots, n-1$.
A Hamiltonian of $C_n$-symmetric system is block diagonalized to be
\begin{align}
  H(k) \to  \mathrm{diag}
  \left(
    H_0(k), \cdots, H_{n-1}(k)
  \right),
\end{align}
where $H_\mu(k) = V^\dag_\mu H(k) V_\mu$, $V_\mu = (\bm v_1, \cdots, \bm v_{g_\mu})$, $ C_n \bm v_j = e^{-i \mu 2\pi /n} \bm v_j$, $C_n V_\mu = V_\mu e^{-i \mu 2\pi / n}$, is Hamiltonian in the $C_n = e^{-i \mu 2\pi /n}$ eigenspace.
$g_\mu$ is the degeneracy of the eigenvalue of $e^{-i \mu 2\pi / n}$ then
$\sum_{\mu=0}^{n-1} g_\mu = \dim H(k)$.

In a superconductor with the $n$-fold rotational symmetry, $[\tilde C_n, H(k)] = 0$, chiral symmetry in the eigenspaces is found when $[\Gamma, \tilde C_n]=0$ holds:
\begin{align}
  \{ \Gamma_\mu, H_\mu(k) \} = 0,
  \
  \Gamma_\mu = V_\mu^\dag \Gamma V_\mu,
  \
  \Gamma_\mu^2 = 1.
\end{align}
Hereafter we assume that the pair potential is the $A$ representation of $C_n$, i.e., $\Delta(\bm k) =  C_n^\dag \Delta(\bm k') C_n$, because $[\Gamma, \tilde C_n] = 0$ holds only in this case.
The winding number in each eigenspace is 
\begin{align}
  W_\mu = \frac{i}{4\pi}
  \int_{-\pi/a_\perp}^{\pi/a_\perp} d k
  \,
  \mathrm{tr}
  \left[
    \Gamma_\mu H_\mu(k)^{-1}
    \frac{\partial H_\mu(k)}{\partial k}
  \right],
\end{align}
which corresponds to the number of zero-energy end states of $H_\mu$.

\subsection{Time-reversal symmetry}

Since the angular momentum is time-reversal odd, one finds
\begin{align}
  \mathcal T^{-1} \tilde C_n \mathcal T = \tilde C_n,
\end{align}
and
\begin{align}
  \mathcal T V_\mu = V_{-\mu} \mathcal T_{-\mu},
  \
  \mathcal T_{\mu} = V^\dag_\mu \mathcal T V_{-\mu}.
\end{align}
These lead to
\begin{align}
  T_{\mu}^\dag H_{\mu}(-k) T_{\mu} = + H_{-\mu}(k)^*,
 \
  T_{\mu}^\dag \Gamma_\mu T_{\mu} = -\Gamma_{-\mu}(k)^*.
\end{align}
As a result, one finds
\begin{align}
 W_\mu &=- \frac{i}{4 \pi} \int_{-\pi}^\pi d k
\,
\mathrm{tr}
\left[
   \Gamma_{-\mu}^* H_{-\mu}(-k)^{* -1} \frac{\partial H_{-\mu}(-k)^*}{\partial k}
\right]
 = -W_{-\mu}^* = -W_{-\mu}.
\label{zeroWmu}
\end{align}
The above relation is a natural extension from Eq. (\ref{zeroW}).

\subsection{Spatial symmetry}

The commutation relation of $C_n$ and spatial symmetries, $U_l$ and $P_l$, is given by
$
 [U_l, C_n] = 0
$ for $[j_z, U_l] = 0$ and $U_l^\dag C_n U_l = C_n^\dag$ for $\{ j_z, U_l \} = 0$.
The same equations hold for $P_l$.
This gives the transformation of Hamiltonian;
\begin{align}
 H_\mu(k) = U_{l, p(j_z, U_l) \mu}^{\dag} H_{p(j_z, U_l) \mu}(k) U_{l, p(j_z, U_l) \mu},
 \
 H_\mu(k) = P_{l, p(j_z, U_l) \mu}^{\dag} H_{p(j_z, U_l) \mu}(-k) P_{l, p(j_z, U_l) \mu},
\end{align}
where $U_{l, \mu} = V^\dag_\mu U_l V_\mu$ and $P_{l, \mu} = V^\dag_\mu P_l V_\mu$.
The chiral operator is transformed by $U_{l, \mu}$ as
\begin{align}
 \Gamma_\mu = \chi(U_l) U_{l, p(j_z, U_l)}^\dag \Gamma_{p(j_z, U_l) \mu} U_{l, p(j_z, U_l)}.
\end{align}
This is the same for $P_{l, \mu}$.
The winding numbers satisfy the following relations;
\begin{align}
 W_\mu = \chi(U_l) W_{p(j_z, U_l) \mu} = -\chi(P_l) W_{p(j_z, P_l) \mu}.
\end{align}
Combining these and Eq. (\ref{zeroWmu}), one finds a necessary condition 
\begin{align}
 \chi(U_l) p(j_z, U_l) =  -\chi(P_l) p(j_z, P_l) = 1,
 \label{zeroWmu2}
\end{align}
for $W_{\mu} \ne 0$.

In symmorphic space groups, from the above conditions,  $W_\mu$ takes a finite value only for the $A_{1u}$ (or its compatible) representation.
The (anti)commutation relations of $j_z$ and the symmetry operations, $U_1 = C_n$, $U_2 = \sigma(x_\perp x_{\parallel 1})$, $U_3 = \sigma(x_\perp x_{\parallel 2})$, $P_1 = \sigma(x_{\parallel 1} x_{\parallel 2})$, $P_2 = C_2(x_{\parallel 1})$, and $P_3 = C_2(x_{\parallel 2})$,  are given by
\begin{align}
&
  [j_z, C_n] 
  = \{ j_z, \sigma(x_\perp x_{\parallel 1}) \} 
  = \{ j_z, \sigma(x_\perp x_{\parallel 2}) \} 
  = [j_z, \sigma(x_{\parallel 1} x_{\parallel 2})] 
  = \{ j_z, C_2(x_{\parallel 1}) \} = \{j_z, C_2(x_{\parallel 2}) \} = 0.
\end{align}
From this and Eq. (\ref{zeroWmu2}), the necessary condition is given by
\begin{align}
 \chi(C_n(x_\perp)) 
 = -\chi(\sigma({x_{\parallel 1} x_{\parallel 2})}) 
 = -\chi(\sigma(x_\perp x_{\parallel 1})) 
 = -\chi(\sigma(x_\perp x_{\parallel 2})) 
 = \chi(C_2(x_{\parallel 1})) 
 = \chi(C_2(x_\perp x_{\parallel 2})) 
 = 1.
\end{align}
This holds for the $A_{1u}$ representation of the pair potential.

\section{Example: Bilayer Rashba system}
\label{examples}



\begin{figure}
\begin{center}
\includegraphics[scale=0.4]{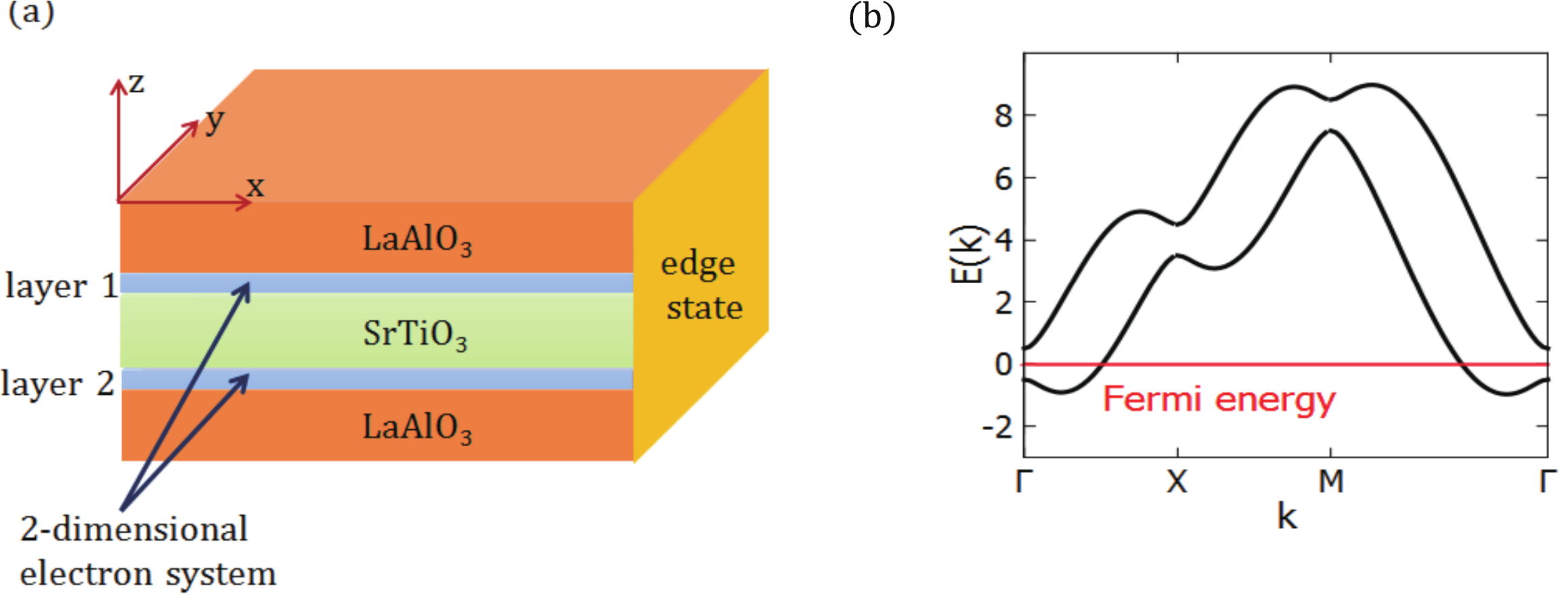}
 \end{center}
\caption{Schematic view of the Rashba bilayer between LaAlO$_{3}$ and SrTiO$_{3}$ \cite{Nakosai12}. 
(a) Two-dimensional electron gases are formed in the interfaces.
The energy dispersions of the finite system  with the edge normal to the $x$ axis are shown in Fig. \ref{fig:one} for possible pair potentials. 
(b) The energy dispersion of the Hamiltonian Eq. (\ref{H0}) in the normal state.  
The Fermi energy is located at $E=0$.
The parameters are taken as $m=0.5$, $\varepsilon =0.5$, $\alpha =2$.}
\label{senko}
\end{figure}

As an example, we show the magnetic response of Majorana Ising spin in the bilayer Rashba superconductor \cite{ Nakosai12}, which are depicted in Fig. \ref{senko}.
The Hamiltonian in the normal state reads
\begin{center}
\begin{align}
{\cal H}_{0}=\frac{k^{2}}{2m} \sigma_0 s_0 - \varepsilon \sigma_{x} s_0 +  \alpha (k_{x}s_{y}-k_{y}s_{x})\sigma_{z},
\
 k = \sqrt{k_x^2 + k_y^2},
 \label{H0}
\end{align}
\end{center}
where $s$ and $\sigma$ denote the Pauli matrices representing the spin and layer degrees of freedom, respectively.
%
%
Time-reversal-invariant Bogoliubov-de Gennes (BdG) Hamiltonian has the form
\begin{eqnarray}
H_{\rm BdG} = 
\begin{pmatrix}
\mathcal H_{0}-\mu & \Delta \\
\Delta & -(\mathcal H_{0} - \mu) 
\end{pmatrix},
\end{eqnarray}
in the basis of
$(c_{\bm k \uparrow}, c_{\bm k \downarrow}, c_{-k\downarrow}^\dag, -c_{-\bm k\uparrow}^\dag)$, where the arrows $\uparrow$ and $\downarrow$ denote the up and down spins, respectively.
When the Fermi level $\mu$ is located within the hybridization gap, as shown in Fig. \ref{senko}(b), the $\mathbb Z_2$ topological invariant takes the nontrivial value \cite{Nakosai12}.
The above Hamiltonian is regularized on the square lattice as
\begin{align}
 \mathcal H_0
 &\to 
 \frac{2-\cos k_x - \cos k_y}{m} \sigma_0 s_0
  - \epsilon \sigma_x s_0
  + \alpha (s_y \sin k_x - s_x \sin k_y) \sigma_z.
\end{align}
We consider the six types of odd-parity pair potentials, which are summarized in Table \ref{pair}.
\begin{table}
\centering
\caption{Irreducible representations $\Gamma$ of odd-parity pair potentials $\Delta$ in the bilayer Rashba superconductor under the $D_{4h}$ symmetry. 
The Majorana Ising spin $\bm S$ on the (100) edge is taken from Table \ref{Wtables}. $\bm 0$ denotes the absence of Majorana fermion.}
\begin{tabular}{ccc}
  \hline\hline
$\Gamma$ & $\Delta$ & $\bm S$
\\ \hline     
 $A_{1u}$ & $\sigma_{y} s_{z}$ & $\bm x$
  \\ 
 $A_{2u}$ & $ \sigma_{z} s_0 $ & $\bm y$
  \\ 
$B_{1u}$ & $\sin k_x \sin k_y \sigma_{z} s_0$ & $\bm x$
  \\ 
$B_{2u}$ & $(\cos k_x - \cos k_y )\sigma_{z} s_0$ & $\bm y$
  \\ 
$E_{u}(x)$ & $\sigma_{y} s_{y}$ & $\bm 0$
 \\ 
$E_{u}(y)$ & $\sigma_{y} s_{x}$ & $\bm z$
\\
  \hline\hline
\end{tabular}
\label{pair}
\end{table}
The corresponding finite-sized Hamiltonian defined in ($1 \leq x \leq N_x$) along the $x$ direction
is given by
\begin{center}
\begin{eqnarray}
	{H}(k_{y})={\cal H}_{0}(k_{y}) +\tilde{\Delta}(k_{y}),
\end{eqnarray}
\end{center}
with
\begin{center}
\begin{eqnarray}
	{H}(k_{y})=\sum_{n=1}^{N_{x}} {c}_{n}^{\dagger} 
\left[  \left( \frac {2 - \cos k_y } {m} + \mu \right ) \sigma_0 s_0
      +\alpha \sin k_{y} \sigma_{z} s_{x} + \varepsilon \sigma_{x} s_0
  \right]
  \tau_z
  {c}_{n} 
  \nonumber\\
  +
  \left[
	\sum_{n=1}^{N_{z}-1}{c}_{n}^{\dagger} \left(  \frac{1}{2m} \sigma_0 s_0 +\frac{\alpha}{2} \sigma_{z}  s_{y}
	\right) 
	\tau_{z} {\tilde c}_{n+1} + \mathrm{h.c.} \right],
\end{eqnarray}
\end{center}
and
\begin{align}
\tilde{\Delta}(k_{y})=\begin{cases}
\displaystyle\sum_{n=1}^{N_{x}} {c}_{n}^{\dagger}  \Delta \sigma_{y} s_{z}\tau_{x} { c}_{n},  
&  \textrm{for} \ A_{1u},
\\[1em]
\displaystyle\sum_{n=1}^{N_{x}} {c}_{n}^{\dagger}  \Delta \sigma_{z} s_{0}\tau_{x} {c}_{n},  
& \textrm{for} \ A_{2u},
\\[1em]
i \displaystyle\sum_{n=1}^{N_{x}-1} {c}_{n+1}^{\dagger} \sin k_{y}  \Delta \sigma_{z} s_{0}\tau_{x} {c}_{n}  
+ \mathrm{h.c.},
  &  \textrm{for} \ B_{1u},
\\[1em]
\displaystyle\sum_{n=1}^{N_{x}} {c}_{n}^{\dagger} \cos k_{y}  \Delta \sigma_{z} s_{0}\tau_{x} {c}_{n}  
-\frac{1}{2}
\left(
\sum_{n=1}^{N_{x}-1} {c}_{n+1}^{\dagger} \Delta \sigma_{z} s_{0}\tau_{x} {c}_{n}   + \mathrm{h.c.}
\right),
 &  \textrm{for} \ B_{2u}, 
\\[1em]
\displaystyle\sum_{n=1}^{N_{x}} {c}_{n}^{\dagger}  \Delta \sigma_{y} s_{y}\tau_{x} {c}_{n},  
&  \textrm{for} \ E_{u}(x) ,
\\[1em]
\displaystyle\sum_{n=1}^{N_{x}} {c}_{n}^{\dagger}  \Delta \sigma_{y} s_{x}\tau_{x} {c}_{n},  
&  \textrm{for} \ E_{u}(y).
\end{cases} 
\end{align}

\begin{figure}
\begin{center}
\includegraphics[scale=0.73]{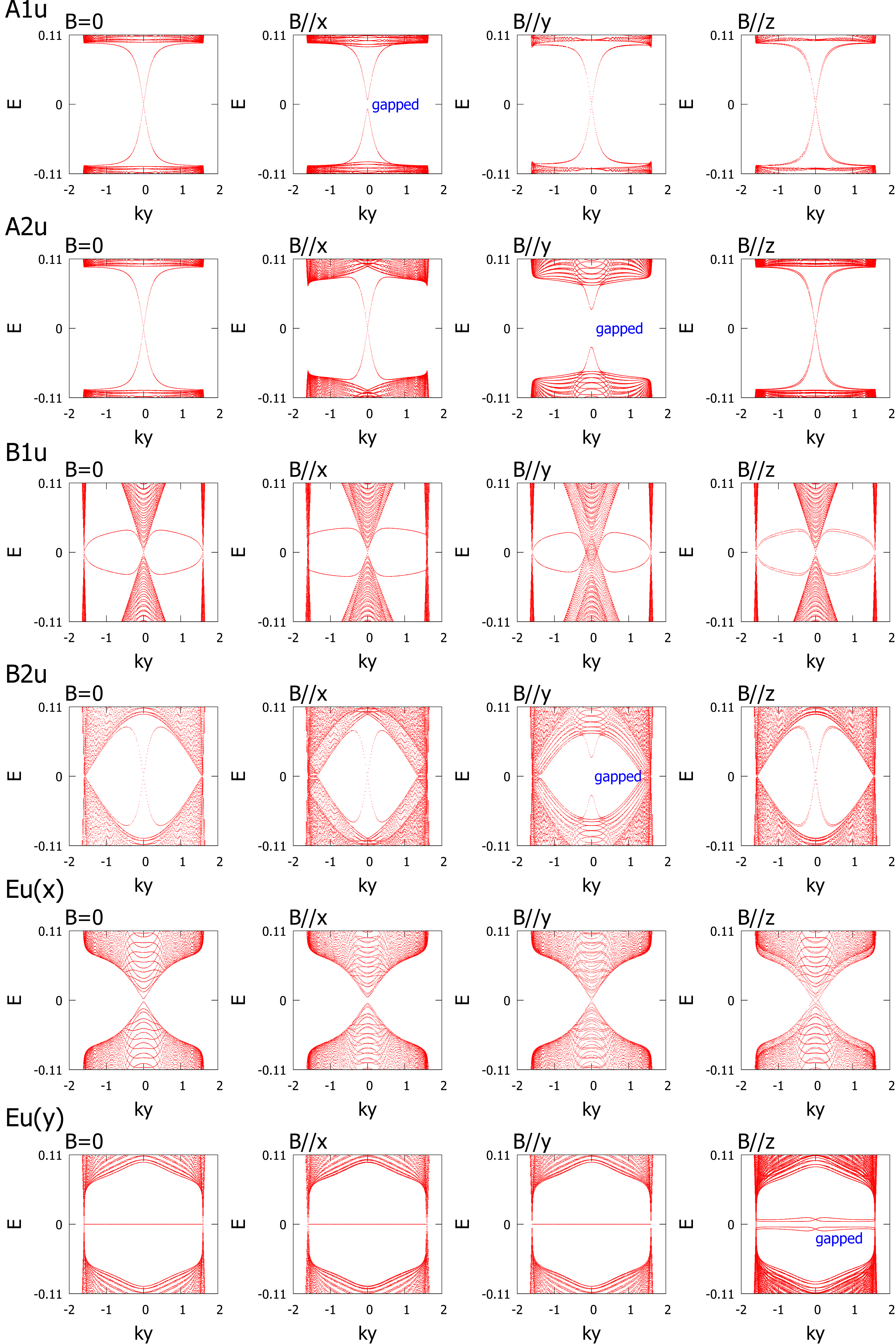}
 \end{center}
\caption{Energy spectra for odd parity pair potentials for $\Delta = 0.1$ and $h=0.03$. The parameters are the same as in Fig. \ref{senko}(b).}
\label{fig:one}
\end{figure}

We calculate the energy spectrum in the presence of a Zeeman field along $x$, $y$, and $z$ directions, which are expressed by the Hamiltonian
\begin{align}
 H_{\bf B} &= \sum_{n=1}^{N_x} c^\dag_n 
 \left( {\bf B}\cdot \frac{g \mu_{\rm B}}{2} {\bf s}
 \right) c_n
=\sum_{n=1}^{N_x} c^\dag_n {\bf h}\cdot {\bf s}
c_n.
\end{align}
For topological superconducting states, the Majorana zero modes still remain gapless in the presence of a magnetic field perpendicular to the Majorana Ising spin.
The direction of the Majorana Ising spin for each pair potential is derived by the general theory studied in the previous section and shown in Table \ref{pair}.
This is verified by the numerical results, which are shown in Fig. \ref{fig:one}.
For the $A_{1u}$ pairing, Majorana zero modes exist at $k_{y}=0$ for the cases of $\bm B \parallel \bm y$ and of $\bm B \parallel \bm z$, while they vanish and a gap is generated for the case of $\bm B \parallel \bm x$.
For the $A_{2u}$ pairing, Majorana zero modes vanish only for the case of $\bm B \parallel \bm y$.
Note that the $A_{2u}$-pairing state under a magnetic field along the $z$ direction is the same as the pair-density-wave (PDW) state studied in Ref \cite{yoshida9, yoshida7}.
There is no Majorana zero mode for the $B_{1u}$ pairing because the bulk superconducting gap closes at $k_y = 0$.
For the $B_{2u}$ pairing, on the other hand, the bulk superconducting gap closes at $k_y \ne 0$ and remains finite at $k_y=0$.
Hence Majorana zero modes are emerged at $k_y=0$ and killed by a magnetic filed along the $y$ direction.
The $E_u$ pairings are similar to the $B_{1u}$ and $B_{2u}$ pairings.
The bulk gap vanishes at $k_y=0$ for the $E_u(x)$ pairing but survives for the $E_u(y)$ paring.
The emerged Majorana zero modes are gapped only when a magnetic field is applied along the $z$ direction.
Namely, $A_{1u}$, $A_{2u}$, $B_{2u}$, and $E_{u}(y)$ pair potentials, Majorana zero modes vanish for a specific direction of magnetic field, i.e., the Majorana zero modes respond to the field as a Ising spin.
This results on the Majorana Ising spins totally coincide with those in Table \ref{pair}.

\section{Conclusion}
\label{conclusion}

We have derived possible nonzero topological invariants and the direction of Majorana Ising spin for each irreducible representation of pair potential for time-reversal-invariant superconductors.
The obtained result is the detailed classification in the class-DIII superconductors in one spatial dimension.
Another point of view is a topological extension to the classification of superconducting pair potential.
Our result is general and does not depend on the detail of systems therefore it is useful for all researchers on superconducting materials.
The anisotropy can be detected by the tunneling spectroscopy under a magnetic field or with a ferromagnetic junction because a zero-bias peak appears in the presence of the Majorana zero modes \cite{kashiwaya00}. 

Several examples have been shown for the bilayer Rashba superconductor, $D_{4h}$, $C_{4v}$, $C_{2v}$ point groups, and the $Pmma$ space group. 
In the other point groups, $D_{6h}$, $D_{3d}$, and $D_{3h}$, another type of anisotropic response can arise.
This will be demonstrated in a separate paper.
As for the topological invariants, we focused only on the $\mathbb Z$ topological invariant in the paper. 
To complete the classification, we also need to clarify the $\mathbb Z_2$ topological invariant and the related Majorana Ising spins.
This issue will be also addressed in a future paper.

\vspace{6pt} 

\acknowledgments{
This work was supported by Grants-in-Aid for Young Scientists (B, Grant No. 16K17725), for Research Activity Start-up (Grand No. JP16H06861), and for Scientific Research on Innovative Areas Topological Material Science JSPS KAKENHI (Grants No. JP15H05851, and No. JP15H05853). 
S.K. was supported by Building of Consortia for the Development of Human Resources in Science and Technology.
}





\appendix

\section{Bulk-edge correspondence in a lattice model}
\label{bec}

In this Appendix, we derive the bulk-edge correspondence, i.e., the number of zero-energy end states equals to the one-dimensional winding number $W$, in lattice models.

\subsection{Number of edge states}

Firstly, we derive the number of zero-energy edge states in the corresponding lattice model with the $r$-th neighbor hoppings, generalizing the discussion by W. Izumida et al. \cite{izumida16}.
The Hamiltonian is expressed in a semi-infinite lattice defined in $n \leq 0$ as
\begin{align}
  H &= \sum_{n \leq 0} 
c^\dag_n \epsilon c_n 
+
 \sum_{q=1}^r \sum_{n \leq -q}
\left(
  c^\dag_n t_q c_{n+q}
 + c_{n+q}^\dag t_q^\dag c_n 
\right),
\
\epsilon = \epsilon_1 \tau_1 + \epsilon_3 \tau_3,
\
t_q = t_{q 1} \tau_1 + t_{q 3} \tau_3.
\end{align}
The Schr\"odinger equation is given by
\begin{align}
 E \psi_n = \epsilon \psi_n + 
\sum_{q=1}^r
\left(
t_q \psi_{n+1} + t_q^\dag \psi_{n-1}
\right),
\end{align}
where $\psi_n$ is the wavefuntion at the $n$-th site.
Now we derive zero-energy ($E=0$) end states, whose fundamental solution of wavefunction has the form 
\begin{align}
  \psi_n = \lambda^{-n} \phi,
\end{align} 
for $|\lambda| < 1$ in the left-half space $n \leq 0$.
Substituting the above form into the Schr\"odinger equation, one obtains
\begin{align}
  \left[\epsilon + 
\sum_{q=1}^r
\left(
t_q \lambda^{-q} + t^\dag_q \lambda^q
\right)
 \right] \phi = 0.
\end{align}
Multiplying $\tau_3$ to the above equation, one finds that the zero-energy states are chirality eigenstates, i.e., $\phi$ is given by $\phi = \phi_\tau \chi_\tau$ for $\tau_2 \chi_\tau =  \tau \chi_\tau$ with $\tau = \pm 1$ chirality.
$\phi_\tau$ is obtain by solving $q_\tau(\lambda) \phi_\tau = 0$ with
\begin{align}
  q_\tau(\lambda) &= \epsilon_1 i \tau + \epsilon_3 
	+ \sum_{q=1}^r 
\left\{
\bigl[t_{q 1} i \tau + t_{q 3} \bigr] \lambda^{-q}
	+ \bigl[t_{q 1}^\dag i \tau + t_{q 3}^\dag \bigr] \lambda^q
\right\},
\end{align}
For a nontrivial solution of $\phi_\tau$, the secular equation
$
 \det q_\tau(\lambda) = 0
$
holds and has [$r \dim H(k)$] solutions.
As a result,
the fundamental solutions for the zero-energy end states are obtained to be
\begin{align}
 \psi_{\tau, n} = 
\lambda_{j}^{-n}
 \phi_\tau(\lambda_j) \chi_\tau,
\
\lambda_j \in Q_\tau =\{\lambda | \det q_\tau(\lambda) = 0, |\lambda|<1 \},
\
q_\tau(\lambda_j) \phi_\tau(\lambda_j) = 0.
\end{align}
The number of the independent solutions for the definite chirality $\tau$ is $|Q_\tau|$.

So as to obtain the physical solutions with a definite chirality $\tau$, a boundary condition is imposed on the system end $n=0$, e.g.,
the fixed boundary condition $\psi_{\tau, q} =  0$ for $q \geq -r + 1$.
The boundary condition gives [$r \dim H(k)/2$] conditions.
Consequently, one obtains the number of the zero-energy end states with the chirality of $\tau$ to be
\begin{align}
 N_\tau = 
 \left|
 |Q_\tau| - \frac{r \dim H(k)}{2}
 \right| 
 \theta \left(|Q_\tau| - \frac{r \dim H(k)}{2} \right).
\end{align}
One obtains only the trivial solution, $\psi_{\tau, n} = 0$,  for $|Q_\tau| = r \dim H(k)/2$.

$\det q_\tau(\lambda) = 0$ is equivalent to $\det q_\tau(\lambda)^* = 0$, which is explicitly shown as
\begin{align}
 \det
\Bigl\{
  \epsilon_1 (-i \tau)
+ \epsilon_3
+ 
\sum_{q=1}^r
\bigl[
  t_{q 1} (-i \tau) + t_{q 3}
\bigr] \lambda^{q *}
+ \sum_{q=1}^r \bigl[
  t_{q 1}^\dag (-i \tau) + t_{q 3}^\dag
\bigr] \lambda^{-q *}
\Bigr\} 
= 0.
\end{align}
This means that if solution is given by $\lambda$ for $\tau = + 1$ then the solution for $\tau = -1$ is given by $\lambda^{* -1}$.
Namely, one finds
\begin{align}
  |Q_+| + |Q_-| = r \dim H(k).
\end{align}
From the above condition, the possible $(N_+, N_-)$ are classified into three cases:
\begin{align}
  (N_+, N_-) = \begin{cases}
  (N, 0), & |Q_+| > \displaystyle\frac{r \dim H(k)}{2},
\\[1em]
  (0, 0), & |Q_+| = \displaystyle\frac{r \dim H(k)}{2},
\\[1em]
  (0, N), & |Q_+| < \displaystyle\frac{r \dim H(k)}{2}.
\end{cases}
\end{align}

\subsection{Bulk-edge correspondence}

Next, we calculate the winding number of the translational-invariant system, which is described by
\begin{align}
 H(k) = \epsilon + \sum_{q=1}^r 
 \left(
 t_q e^{i k q} + t_q^\dag e^{-i k q}
 \right).
\end{align}
The winding number is given by
\begin{align}
 W &= 
  \frac{i}{4 \pi}\int_{-\pi}^\pi
 dk \mathrm{tr}
 \left[
   \Gamma H(k)^{-1} \frac{\partial H(k)}{\partial k}
 \right]
 \nonumber\\ &
 =
 \frac{i}{4 \pi} \int_{-\pi}^\pi d k \, \mathrm{tr}
 \Biggl[
	\begin{pmatrix}
		1 & 0
	\\
		0 & -1
	\end{pmatrix}
	\begin{pmatrix}
		0 & q^{\dag -1}(k)
\\
	q^{-1}(k) & 0
	\end{pmatrix}
\begin{pmatrix}
	0 & \displaystyle\frac{\partial q(k)}{\partial k}
\\
 \displaystyle\frac{\partial q(k)^\dag}{\partial k} & 0
\end{pmatrix}
\Biggr]
\nonumber\\
 &= \mathrm{Im} \int_{-\pi}^\pi \frac{d k}{2 \pi} \frac{\partial}{\partial k} \ln \det q(k).
\end{align}
In this way, the basis in which the chiral operator is diagonalized makes it easy to calculate the winding number.
Then we first 
off-diagonalize the Hamiltonian as
\begin{align}
 \tilde H(k) = \Lambda_3(k) \tau_1 + \Lambda_1(k) \tau_2,
 \
  \Lambda_j(k) = \epsilon_j + \sum_{q=1}^r
\left(
 t_{q j} e^{i k q} + t_{q j}^\dag e^{-i k q}
\right).
\end{align}
This reduces the winding number to be
\begin{align}
  W &= 
\mathrm{Im} \int_{-\pi}^\pi \frac{d k}{2\pi} \frac{\partial}{\partial k} \ln \det \left[ \Lambda_3(k) - i \Lambda_1(k) \right]
 = \mathrm{Im} \int_{-\pi}^\pi \frac{d k}{2\pi} \frac{\partial}{\partial k} \ln \det q_{-} (e^{-i k})
\nonumber\\ & 
 = -\mathrm{Im} \oint_{|\lambda|=1} \frac{d \lambda}{2 \pi} \frac{\partial}{\partial \lambda} \ln \det q_-(\lambda)
 = -|Q_-| + \frac{r}{2} \dim H(k),
\end{align}
where the integral runs over the unit circle in the complex $\lambda$ plane along the counter-clockwise direction.
The last line of the above equation is derived with the use of the argument principle because $\det q_\tau(\lambda)$ has $|Q_\tau|$ zeros within the unit circle and 
 is asymptotically given by
\begin{align}
 \det q_\tau(\lambda)
 \sim 
 \frac{1}{\lambda^{r [\dim H(k)]/2}}
 \det (t_{r1} i \tau + t_{r3}),
\end{align}
which has the order-$[r \dim H(k)]/2$ pole.
Thus $W = -N_-$ for $|Q_-| > [r \dim H(k)]/2$.
Because $W$ is an integer, the above relation is rewritten as
\begin{align}
  W &= W^* = - \mathrm{Im} 
\int_{-\pi}^\pi \frac{d k}{2\pi} \frac{\partial}{\partial k} \ln \det \left[ \Lambda_3(k) + i \Lambda_1(k) \right]
 = |Q_+| - \frac{r}{2} \dim H(k).
\end{align}
The winding number equals $W = N_+$ for $|Q_+| > [r \dim H(k)]/2$.
We finally arrive at the bulk-edge correspondence:
\begin{align}
 W = N_+ - N_-.
\end{align}

\section{Representation of symmetry operation}
\label{rep}

Order-2 symmetry operations of space groups are decomposed into the real-space part $O(\bm \tau)$ and spin part $\Sigma$;
\begin{align}
 \{ U | \bm \tau \} = O(\bm \tau) \Sigma,
\end{align}
where $\bm \tau$ is half a primitive translation vector.
The spin part is independent of the translation.
For (screw) rotations, $\Sigma$ is taken to be along the (screw) rotational axis.
For (glide) reflections, on the other hand, $\Sigma$ is perpendicular to the (glide) reflection plane, $\Sigma = \bm s \cdot \bm n$, where $s_x$, $s_y$, $s_z$ are the Pauli matrices and $\bm n$ is the unit vector normal to the plane.
Note that $\Sigma^2 = 1$ and $\Sigma$ anticommutes with time reversal, $\{ \Sigma, \mathcal T \} = 0$.
The real-space part $O(\bm \tau)$ commutes with time reversal $[O(\bm \tau), \mathcal T] = 0$ and satisfies the following relation
\begin{align}
 O(\bm \tau)^2 = \{1 | 2 \bm \tau\} = e^{i 2 \bm k \cdot \bm \tau},
\end{align}
in the momentum space.
Therefore, the chiral operator is given by
\begin{align}
 \Gamma_{\{ U | \bm \tau \}} = e^{-i \bm k \cdot \bm \tau} \{ U | \bm \tau \} \tau_y,
 \
 \Gamma_{\{ U | \bm \tau \}}^2 = 1.
\end{align}
In this choice, we obtain the commutation relation of $\Gamma_{\{ U | \bm \tau \}}$ and time reversal $\mathcal T$;
\begin{align}
 \{ \mathcal T, \Gamma_{\{ U | \bm 0 \} } \} &= 0,
 \\
 \{ \mathcal T, \Gamma_{\{ U | \bm \tau \} } \} &= 0, \textrm{ for } e^{i \bm k \cdot \bm \tau} = 1,
 \\
 [ \mathcal T, \Gamma_{\{ U | \bm \tau \} } ] &= 0, \textrm{ for } e^{i \bm k \cdot \bm \tau} = \pm i.
\end{align}
For nonsymmorphic symmetry operations, $\bm \tau \ne \bm 0$, representations at the zone center are different from those at the zone boundary.

\section{Tables for irreducible representations and Majorana Ising spins}
\label{tables}

We show several examples of topological invariants $W[\tilde U_i, \bm x_\perp]$ under the $D_{4h}$, $C_{4v}$, and $C_{2v}$ point-group symmetries in Table \ref{Wtables}.
The nonzero topological invariants, $W[\tilde U_i, \bm x_\perp]$, are derived by applying Eq. (\ref{chis}) to each representation.
We also show the case for the space group $Pmma$ that is
\begin{align}
 \left\{ 
 	\{ C_2(z) | \bm a/2 \},
 	\{ C_2(x) | \bm a/2 \},
 	\{ C_2(y) | \bm 0 \},
 	\{ \sigma(xy) | \bm a/2 \},
  	\{ \sigma(yz) | \bm a/2 \},
  	\{ \sigma(xz) | \bm 0 \}
 \right\},
\end{align}
 in Table \ref{Wtables2}, 
solving Eq. (\ref{condition}) and the commutators, $p(\tilde U_j, \tilde U_l)$, which are defined by $U_j U_l = p(U_j, U_l) U_l U_j$, for the symmetry operations $U_j$ in the $Pmma$ group (Table \ref{pjl}). 
$p(U_j, U_l)$ is calculated in the same manner as in Ref. \cite{kobayashi16}.
Note that the topological invariants for glide reflections vanish, $W[\{ U | \bm a /2 \}, \bm x_\perp] = 0$, for $k_x a_x = \pi$, as discussed in Sec. \ref{ti}.
And also, Majorana fermions do not appear on the $(yz)$ plane even for $W[\{ U | \bm a /2 \}, \bm x] \ne 0$ since the nonsymmorphic symmetry involving the half translation of $\bm a / 2$ is broken on the $(yz)$ surface.
Finally, the Majorana Ising spin protected by $W[\tilde U_i, \bm x_\perp]$ is obtained to be parallel to the rotational axis for $\tilde U_i = \tilde C_2$ or normal to the mirror plane for $\tilde U_i = \sigma$, respectively, as explained in Sec. \ref{preliminary}.

\begin{table*}
\centering
\caption{Possible topological invariants $W[{\tilde U_i, \bm x_\perp}]$ and the direction of the Majorana Ising spin on the given surfaces for each irreducible representation of $D_{4h}$, $C_{4v}$, and $C_{2v}$.
The surface is denoted by the mirror index $(hkl)$ or the Cartesian coordinate $(x_i x_j)$.
$\bm x_\sigma$ of $W[\sigma, \bm x_{\sigma}]$ is the direction normal to the surface and   is on the $\sigma$ mirror plane.
The $C_2'$ axis is set to the $x$ axis.
The representations $E_u(x)$ and $E_u(y)$ are defined in the bottom table.
}
\label{Wtables}
\begin{tabular}{ccccc}
\hline\hline
$D_{4h}$ & $W[{U, \bm x_\perp}]$ & $(001)$ & $(100)$ & $(110)$
\\
\hline
$A_{1u}$ & $W[{C_{n}}]$, $W[{C_2'}]$, $W[{C_2''}]$ & $[001]$ & $[100]$ & $[110]$
\\
$A_{2u}$ & $W[{\sigma_v, \bm x_{\sigma_v} \ne [001]}]$, $W[{\sigma_d, \bm x_{\sigma_d} \ne [001]}]$ & $\bm 0$ & $[010]$ & $[1 \bar 1 0]$
\\
$B_{1u}$ & $W[{C_2'}]$, $W[{\sigma_d, \bm x_{\sigma_d} \ne [001]}]$ & $\bm 0$ & $[100]$ & $[1 \bar 1 0]$
\\
$B_{2u}$ & $W[{C_2''}]$, $W[{\sigma_v, \bm x_{\sigma_v} \ne [001]}]$ & $\bm 0$ & $[0 1 0]$ & $[1 1 0 ]$
\\
\hline\hline
\end{tabular}
\\[1em]
\begin{tabular}{cccccc}
\hline\hline
$D_{4h}$ & $W[{U, \bm x_\perp}]$ & $(001)$ & (100) & $(010)$ & (110)
 \\
 \hline
 $E_{u}(x)$ & $W[{\sigma_v(010), \bm x_{\sigma_v(010)} \ne [100]}]$, $W[{\sigma_h, \bm x_{\sigma_h} \ne [100]}]$ & $[010]$ & $\bm 0$ & $[001]$ & $[001]$
 \\
 $E_u(y)$ & $W[{\sigma_v(100), \bm x_{\sigma_v(100)} \ne [010]}]$, $W[{\sigma_h, \bm x_{\sigma_h} \ne [010]}]$ & $[100]$ & $[001]$ & $\bm 0$ & $[001]$
 \\
 \hline\hline
\end{tabular}
\\[1em]
\begin{tabular}{ccccc}
 \hline\hline
 $C_{4v}$ & $W[U, \bm x_\perp]$ & $(001)$ & $(100)$ & (110)
 \\
 \hline
 $A_1$ & $W[{\sigma_v, \bm x_{\sigma_v} \ne \bm z}]$, 
 $W[{\sigma_d, \bm x_{\sigma_d} \ne \bm z}]$ 
 & $\bm 0$ & $[010]$ & $[1 \bar 1 0]$
 \\
 $A_2$ & $W[{C_2}]$ & $[001]$ & $\bm 0$ & $\bm 0$
 \\
 $B_1$ & $W[{\sigma_v, \bm x_{\sigma_v} \ne \bm z}]$ & $\bm 0$ & $[010]$ & $\bm 0$
 \\
 $B_2$ & $W[{\sigma_d, \bm x_{\sigma_d} \ne \bm z}]$ & $\bm 0$ & $\bm 0$ & $[1 \bar 1 0]$
 \\
 $E(x)$ & $W[\sigma_v(zx), \bm x_{\sigma_v(zx)} \ne \bm x]$ 
 & $[010]$ & $\bm 0$ & $\bm 0$
 \\
 $E(y)$ & $W[{\sigma_v(yz), \bm x_{\sigma_v(yz)} \ne \bm y}]$ 
 & $[100]$ & $\bm 0$ & $\bm 0$
 \\
 \hline\hline
\end{tabular}
\\[1em]
\begin{tabular}{ccccc}
 \hline\hline
 $C_{2v}$ & $W[U, \bm x_\perp]$ & $(xy)$ & $(yz)$ & $(zx)$
  \\
  \hline
  $A_1$ & $W[{\sigma_v(zx), \bm x_{\sigma_v(zx)} \ne \bm z}]$, 
  $W[{\sigma_v(yz), \bm x_{\sigma_v(yz)} \ne \bm z}]$ 
  & $\bm 0$ & $\bm y$ & $\bm x$
  \\
  $A_2$ & $W[{C_2}]$ & $\bm z$ & $\bm 0$ & $\bm 0$
  \\
  $B_1$ & $W[{\sigma_v(zx), \bm x_{\sigma_v(zx)} \ne \bm x}]$ & $\bm y$ & $\bm 0$ & $\bm 0$
  \\
  $B_2$ & $W[{\sigma_v(yz), \bm x_{\sigma_v(yz)} \ne \bm y}]$ & $\bm x$ & $\bm 0$ & $\bm 0$
  \\
  \hline\hline
\end{tabular}
\\[1em]
\begin{tabular}{ccccccc}
 \hline\hline
 $D_{4h}$ & $C_2$ & $C_2'(x)$ & $C_2'(y)$ & $\sigma_h$ & $\sigma_v(yz)$ & $\sigma_v(xz)$
 \\ \hline
 $E_u(x)$ & $-$ & $+$ & $-$ & $+$ & $-$ & $+$
 \\
 $E_u(y)$ & $-$ & $-$ & $+$ & $+$ & $+$ & $-$
 \\
 \hline\hline
\end{tabular}
\label{tab1}
\end{table*}

\begin{table}
\centering
\caption{Possible topological invariants $W[\tilde U_i, \bm x_\perp]$ in the $Pmma$ space group for $k_x a_x = 0$ (upper) and $k_x a_x = \pi$ (lower).}
\begin{tabular}{ccccc}
\hline\hline
$Pmma$ & $W[U, \bm x_\perp] (k_x = 0)$ & $(xy)$ & $(yz)$ & $(xz)$
\\
\hline
$A_g$ & 0 & $\bm 0$ & $\bm 0$ & $\bm 0$
\\
$B_{1g}$ & 0 & $\bm 0$ & $\bm 0$ & $\bm 0$
\\
$B_{2g}$ & 0 & $\bm 0$ & $\bm 0$ & $\bm 0$
\\
$B_{3g}$ & 0 & $\bm 0$ & $\bm 0$ & $\bm 0$
\\
$A_u$ & $W[\{ C_2(z) | \bm a/2 \}]$, $W[\{ C_2(x) | \bm a/2 \}]$, $W[\{ C_2(y) | \bm 0 \}]$ & $\bm z$ & $\bm 0$ & $\bm x$
\\
 & $W[\{ \sigma(xy) | \bm a / 2 \}, \bm x_{\sigma(xy)} \ne \bm x, \bm y \}]$
 \\ &
 $W[\{ \sigma(yz) | \bm a / 2 \}, \bm x_{\sigma(yz)} \ne \bm y, \bm z \}]$
 \\ &
 $W[\{ \sigma(xz) | \bm 0 \}, \bm x_{\sigma(xz)} \ne \bm x, \bm z \}]$
\\
$B_{1u}$ 
	& $W[\{ \sigma(xy) | \bm a / 2 \}, \bm x_{\sigma(xy)} \ne \bm x, \bm y \}]$
	& $\bm 0$ & $\bm y$ & $\bm x$
	\\
	& $W[\{ \sigma(yz) | \bm a / 2 \}, \bm x_{\sigma(yz)} \ne \bm z \}]$
	\\
	& $W[\{ \sigma(xz) | \bm 0 \}, \bm x_{\sigma(xz)} \ne \bm z \}]$
\\
$B_{2u}$ 
	& $W[\{ \sigma(xy) | \bm a / 2 \}, \bm x_{\sigma(xy)} \ne \bm y \}]$
	& $\bm x$ & $\bm 0$ & $\bm 0$
	\\
	& $W[\{ \sigma(yz) | \bm a / 2 \}, \bm x_{\sigma(yz)} \ne \bm y \}]$
	\\
	& $W[\{ \sigma(xz) | \bm 0 \}, \bm x_{\sigma(xz)} \ne \bm x, \bm z \}]$
\\
$B_{3u}$
	& $W[\{ \sigma(xy) | \bm a / 2 \}, \bm x_{\sigma(xy)} \ne \bm x \}]$
	& $\bm y$ & $\bm 0$ & $\bm z$
	\\
	& $W[\{ \sigma(yz) | \bm a / 2 \}, \bm x_{\sigma(yz)} \ne \bm y, \bm z \}]$
	\\
	& $W[\{ \sigma(xz) | \bm 0 \}, \bm x_{\sigma(xz)} \ne \bm x \}]$
\\
\hline\hline
\\
\hline\hline
$Pmma$ & $W[U, \bm x_\perp] (k_x a_x = \pi)$ & $(xy)$ & $(yz)$ & $(xz)$
\\
\hline
$A_g$ 
	& 
	$W[\{\sigma(yz) | \bm a/2\}, \bm x_{\sigma(yz)} \ne \bm z]$
	& $\bm 0$ & $\bm 0$ & $\bm x$
\\
$B_{1g}$ &
	$W[\{C_2(z) | \bm a/2 \}]$ & $\bm z$ & $\bm 0$ & $\bm 0$
	\\ &
	$W[\{\sigma(xy) | \bm a/2\}, \bm x_{\sigma(xy)} \ne \bm y]$
	\\ &
	$W[\{\sigma(yz) | \bm a/2\}, \bm x_{\sigma(yz)} \ne \bm y, \bm z]$
\\
$B_{2g}$ &
	$W[\{\sigma(xy) | \bm a/2\}, \bm x_{\sigma(xy)} \ne \bm x, \bm y]$ & $\bm 0$ & $\bm 0$ & $\bm 0$
	\\ &
	$W[\{\sigma(yz) | \bm a/2\}, \bm x_{\sigma(yz)} \ne \bm y, \bm z]$
\\
$B_{3g}$ &
	$W[\{\sigma(xy) | \bm a/2\}, \bm x_{\sigma(xy)} \ne \bm x, \bm y]$ & $\bm x$ & $\bm 0$ & $\bm 0$
	\\ &
	$W[\{\sigma(yz) | \bm a/2\}, \bm x_{\sigma(yz)} \ne \bm y]$
\\
$A_u$ & 
	$W[\{\sigma(xz) | \bm 0 \}, \bm x_{\sigma(xz)} \ne \bm x, \bm z]$ & $\bm 0$ & $\bm 0$ & $\bm 0$
\\
$B_{1u}$ &
	$W[\{\sigma(xz) | \bm 0\}, \bm x_{\sigma(xz)} \ne \bm z]$ & $\bm 0$ & $\bm y$ & $\bm 0$
\\
$B_{2u}$ &
	$W[\{C_2(y) | \bm 0 \}]$,
	$W[\{\sigma(xz) | \bm 0 \}, \bm x_{\sigma(xz)} \ne \bm x, \bm z]$ & $\bm 0$ & $\bm 0$ & $\bm y$
\\
$B_{3u}$ &
	$W[\{C_2(x) | \bm a/2\}]$,
	$W[\{\sigma(xz) | \bm 0 \}, \bm x_{\sigma(xz)} \ne \bm x]$
	& $\bm y$ & $\bm 0$ & $\bm 0$
\\
\hline\hline
\end{tabular}
\label{Wtables2}
\end{table}

\begin{table}
\centering
\caption{Commutator $p(U_j, U_l) = U_j^{-1} U_l^{-1} U_j U_l$ for the $Pmma$ group. A row and column correspond to $U_j$ and $U_l$, respectively.}
\begin{tabular}{c|cccccc}
\hline\hline
 & $\{ C_2(z) | \bm a / 2 \}$ & $\{ C_2(x) | \bm a / 2 \}$ & $\{ C_2(y) | \bm 0 \}$ & $\{ \sigma(xy) | \bm a / 2 \}$ & $\{ \sigma(yz) | \bm a / 2 \}$ & $\{ \sigma(xz) | \bm 0 \}$
\\ \hline
$\{ C_2(z) | \bm a / 2 \}$ & 1 & $-e^{i k_x a_x}$ & $- e^{i k_x a_x}$ & $e^{i k_x a_x}$ & $-1$ & $-1$
\\
$\{ C_2(x) | \bm a / 2 \}$ & $-e^{-i k_x a_x}$ & 1 & $-e^{-ik_x a_x}$ & $-1$ & $e^{ik_x a_x}$ & $-1$
\\
$\{ C_2(y) | \bm 0 \}$ & $-e^{-i k_x a_x}$ & $-e^{i k_x a_x}$ & 1 & $-e^{i k_x a_x}$ & $- e^{i k_x a_x}$ & $1$
\\
$\{ \sigma(xy) | \bm a / 2 \}$ & $e^{-i k_x a_x}$ & $-1$ & $-e^{-i k_x a_x}$ & 1 & $-e^{-i k_x a_x}$ & $-1$
\\
$\{ \sigma(yz) | \bm a / 2 \}$ & $-1$ & $e^{-i k_x a_x}$ & $-e^{-ik_xa_x}$ & $-e^{i k_x a_x}$ & 1 & $-1$
\\
$\{ \sigma(xz) | \bm 0 \}$ & $-1$ & $-1$ & 1 & $-1$ & $-1$ & 1
\\
\hline\hline
\end{tabular}
\label{pjl}
\end{table}

\bibliography{majoranaising}

\end{document}